\documentclass[aip,jap,reprint]{revtex4-1}

\usepackage{graphicx}
\usepackage{lmodern}
\usepackage{multirow}

\usepackage[colorlinks]{hyperref}
\hypersetup{
   pdftitle={Intrinsic and doped coupled quantum dots created by local modulation of implantation in a silicon nanowire},
   pdfauthor={M. Pierre, B. Roche, R. Wacquez, X. Jehl, M. Sanquer and M. Vinet},
   pdfkeywords={carrier density, elemental semiconductors, MOSFET, semiconductor doping, semiconductor quantum dots, silicon},
   urlcolor=blue,
   citecolor=blue,
   linkcolor=blue
   }

\begin{document}

\title{Intrinsic and doped coupled quantum dots created by local modulation of implantation in a silicon nanowire}

\author{M. Pierre}
\author{B. Roche}
\author{R. Wacquez}
\author{X. Jehl}
\author{M. Sanquer}
\email[]{marc.sanquer@cea.fr}
\affiliation{CEA-DSM-INAC-SPSMS, 38054 Grenoble, France}
\author{M. Vinet}
\affiliation{CEA/LETI-MINATEC, 38054 Grenoble, France}
\date{\today}

\begin{abstract}
We present a systematic study of various ways (top gates, local doping, substrate bias) to fabricate and tune multi-dot structures in silicon nanowire multigate MOSFETs (metal-oxide-semiconductor field-effect transistors). The carrier concentration profile of the silicon nanowire is a key parameter to control the formation of tunnel barriers and single-electron islands. It is determined both by the doping profile of the nanowire and by the voltages applied to the top gates and to the substrate. Local doping is achieved with the realisation of up to two arsenic implantation steps in combination with gates and nitride spacers acting as a mask. We compare nominally identical devices with different implantations and different voltages applied to the substrate, leading to the realisation of both intrinsic and doped coupled dot structures. We demonstrate devices in which all the tunnel resistances towards the electrodes and between the dots can be independently tuned with the control top gates wrapping the silicon nanowire.
\end{abstract}
\maketitle

\section{INTRODUCTION}

Most of the  silicon quantum dots fabricated to date are either based on uniformly doped \cite{Scott-Thomas1989, Matsuoka1994, Leobandung1995, Ishikuro1997, Augke2000, Emiroglu2003,Huang2008,Manoharan2008,Tracy2010} or intrinsic \cite{Fujiwara2006,Angus2007} silicon or Si/SiGe heterostructure \cite{Thalakulam2010}. On the contrary, silicon MOSFETs (Metal Oxide Semiconductor Field Effect Transistors) rely on a strong doping gradient between the body and the highly doped source and drain. In today's submicrometer transistors, this non-uniform doping profile is usually realized by two successive steps of self-aligned ion implantation. 
This scheme yields a good trade-off between access resistance and dopant diffusion into the channel.
To achieve this  profile, silicon nitride spacers are formed around the gate. The second implantation step, named Highly Doped Drain (HDD) implantation, is performed after the deposition of these spacers which act as blanket masks. HDD regions are therefore shifted away from the channel and this prevents massive dopants diffusion into the channel. 
To reduce the access resistance a first and moderate implantation step, named Lighly Doped Drain (LDD) implantation, is performed after etching of the gate but before the deposition of the spacers. 
Without LDD implantation the extensions below the spacers remains intrinsic and their resistance is fairly high. For narrow geometries, such as silicon-on-insulator (SOI) nanowire MOSFETs, it can be larger than the quantum of resistance. This enables to build a MOS-SET (Metal-Oxide-Semiconductor Single Electron Transistor) exhibiting Coulomb blockade oscillations at low temperature \cite{Hofheinz06B}.

In this work we demonstrate that LDD implantation is the crucial parameter which controls the crossover between nanowire MOSFETs and MOS-SETs in single gate devices (section~\ref{sec:LDD}). We use  LDD implantation together with multigate designs to build compact and fully tunable highly doped quantum dots systems (section~\ref{sec:singledotLDD}).
In addition we compare the effect of LDD with a substrate bias \cite{Eminente2007}, which is another control parameter for silicon SETs (section~\ref{sec:substrate_bias}). 
 
The fabrication of single and coupled (section~\ref{sec:tunable_double_dot}) highly doped Silicon SETs extends our previous work, which was limited to the fabrication of intrinsic single or coupled MOS-SETs.
Intrinsic coupled MOS-SETs can indeed be obtained without LDD \cite{Pierre2009} or by using two layers of gates, an upper and several top gates \cite{Fujiwara2006,Liu08}. 
Nevertheless the possibility to create doped  MOS-SETs and to combine them with intrinsic MOS-SETs has extra advantages. First highly doped SETs with a large density of carriers
(over $8\times10^{18}\,\mathrm{cm}^{-3}$) have a metallic-like behaviour, the Thomas-Fermi length being comparable to inter-dopant distance. This enables strong screening of the disorder potential, without the need of strong electric fields applied by gates. 
Doped SETs also present sharp and fixed confinement potential. 
Moreover local implantation is particularly attractive in view of the single dopant limit \cite{Ono2007,Tan2010,Morello2009,Pierre2010,Hofheinz06,Lansbergen2008,Schenkel03}. By adapting the dose in standard CMOS implantation one can get on average one dopant per device. A first step toward this goal has been achieved recently \cite{Johnson2010} in similar devices. At this limit there are striking differences between the confinement potential created by a single ionized donor and a transverse electric field \cite{Lansbergen2008}.

Finally, the coupled silicon dots presented here use only SOI MOSFET standard technology, whereas previously reported devices use hybrid metallic and silicon SETs \cite{Buehler2005,Mitic2006,Chan2006} or unconventional gate stack \cite{Fujiwara2006,Liu08}. In addition 2D electron gases were used for source and drain instead of bulk low resistive HDD contacts.

\begin{figure*}[!p]
\begin{center}
\includegraphics[width=\textwidth]{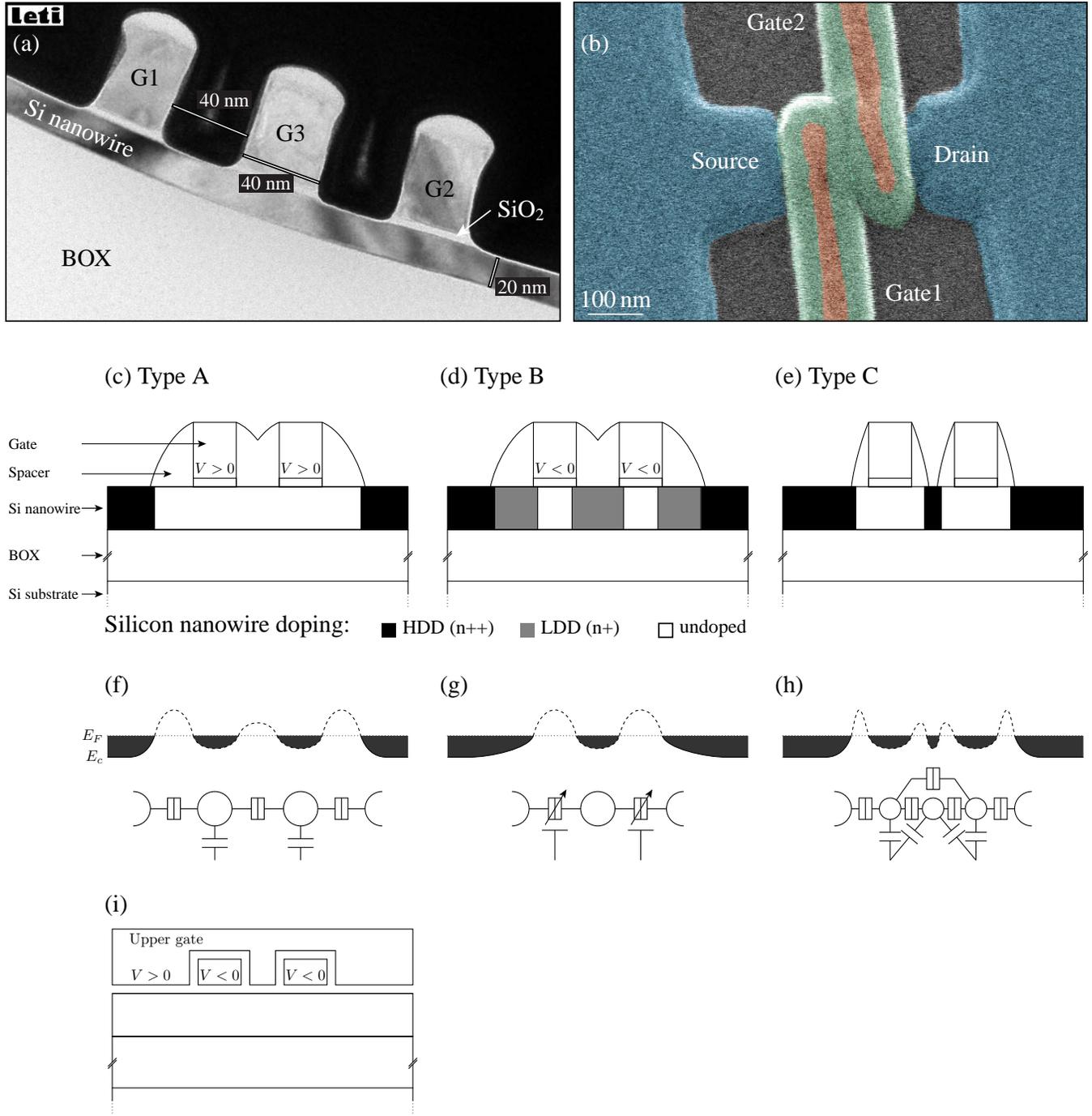}
\caption{(a) TEM cross-view of a typical triple gate sample before spacer deposition. The gate length is 40\,nm and the spacing between gates is 40\,nm. (b) SEM top-view of a typical double gate sample after spacers (in green) are formed all around the gates (in red). The silicon nanowire (in blue) is completely covered from source to drain. (c,d,e,i) Schematic cross-views of different kinds of two-gate samples. The doping of the silicon nanowire is indicated with grey levels. $V<0$ indicates depletion gates used to create tunnel barriers. $V>0$ indicates gates used to accumulate electrons. (c) A-type: without LDD implantation. (d) B-type: with LDD implantation. (e) C-type: with small spacers. (f,g,h) Longitudinal band structures and equivalent single-electron circuits corresponding to (c), (d) and (e) when no voltage is applied on the substrate. (f) A double dot is created by accumulation of carriers below the two gates (see Ref.~\onlinecite{Pierre2009}). (g) A dot is created by implantation of Arsenic dopants between the gates. Its coupling to the electrodes is tuned by the two top gates. (h) A triple dot is created \cite{Pierre2009} both by accumulation of carriers below the gates and implantation of arsenic dopants between them. (i) Alternative design with two levels of gates \cite{Fujiwara2006,Angus2007} (an upper and two top gates) and no implantation of the silicon nanowire.}
\label{fig1}
\end{center}
\end{figure*}

\begin{table*}[!]
 \centering
 \begin{tabular}{|c||c|c|c|c|c|c|c|c|}
  \hline
  sample & number of gates & $L_g$ (nm)  & $W$ (nm)  & $T_\mathrm{Si}$ (nm) & $L_{gg}$ (nm) &  spacers length (nm) & LDD & type \\ \hline \hline
  1 & \multirow{6}{*}{2} & 60 & 60  & 20 & 40 & 40 & No  & A \\ \cline{1-1}\cline{3-9}
  2 &                    & 60 & 60  & 20 & 40 & 40 & Yes & B \\ \cline{1-1}\cline{3-9}
  3 &                    & 50 & 60  & 20 & 50 & 40 & No  & A \\ \cline{1-1}\cline{3-9}
  4 &                    & 50 & 60  & 20 & 50 & 40 & Yes & B \\ \cline{1-1}\cline{3-9}
  5 &                    & 50 & 60  & 20 & 50 & 40 & No  & A \\ \cline{1-1}\cline{3-9}
  6 &                    & 40 & 100 & 25 & 50 & 15 & No  & C \\ \hline                
  7 & 3                  & 40 & 60  & 20 & 50 & 40 & Yes & B \\ \hline                 
\end{tabular}
\caption{Properties of the samples compared in the article. The spacer length and whether LDD is performed determine the sample type. Samples \{1, 2\} and \{3, 4, 5\} are nominally identical apart from LDD implantation.}
\label{tab:samples}
\end{table*}

\section{SAMPLES}
\label{sec:samples}

Our samples are produced on 200\,mm silicon-on-insulator (SOI) wafers in a CMOS platform. Thousands of samples have been built and measured. All devices cooled down to 4.2\,K behave as single-electron devices and nominally identical devices behave the same way with respect to charging energy and Coulomb blockade oscillations periods. A 200 to 500\,nm long, 20\,nm thick and 60 to 100\,nm wide silicon nanowire is made by electron-beam lithography and reactive ion etching and covered by two or three $L_g = 40$--60\,nm long polysilicon gates which are isolated  from silicon by 5\,nm thick SiO$_\mathrm{2}$ gate oxide (Fig.~\ref{fig1}a and Fig.~\ref{fig1}b).
The spacing between gates is $L_{gg} = 40$ or 50\,nm. 15\,nm or 40\,nm thick silicon nitride spacers are formed on both sides of the gates.
LDD doping is optionally performed after gate and before  spacer formation. Without LDD the samples are those already presented in Ref.~\onlinecite{Pierre2009}.
The LDD implantation dose  is $5\times10^{13}\,\mathrm{As}.\mathrm{cm}^{-2}$. This corresponds to a doping level of the SOI of $2\times10^{19}\,\mathrm{cm}^{-3}$ (estimation made using the Crystal TRIM simulator).
The number of As atoms implanted in the central region between the gates is approximately $10^3$ ($\approx 2\times10^{19}\,\mathrm{cm}^{-3} \times 40\,\mathrm{nm} \times 60\,\mathrm{nm} \times 20\,\mathrm{nm}$).
HDD implantation is performed after spacer deposition  using a dose of $2\times10^{15}\,\mathrm{As}.\mathrm{cm}^{-2}$, giving an As concentration larger than $10^{20}\,\mathrm{cm}^{-3}$.
The silicon nanowire lies on the 145\,nm thick buried oxide layer of the SOI wafer. A voltage can be applied to the  silicon substrate, thus acting as a backgate.

Fig.~\ref{fig1} shows the different systems compared in this article. They are depicted with two gates, while an extension to three-gate devices is considered at the end of the article. On Fig.~\ref{fig1}c the nanowire is covered from source to drain by either the gates or the spacers (A-type samples). As a result it is undoped. At low temperature, as far as no substrate bias is applied, intrinsic quantum dots are formed below the gates by locally attracting  electrons with positive gate voltages \cite{Hofheinz06B,Pierre2009}. On Fig.~\ref{fig1}d LDD doping has been done before the spacers deposition (B-type samples). In this case the regions below the gates act as tunable tunnel barriers and electrons can be confined between two gates, forming a doped quantum dot. The conductance of these tunable tunnel barriers is controlled by field effect and is nonzero for gate voltages above threshold. 
Fig.~\ref{fig1}e describes samples with small spacers (15 nm) without LDD (C-type samples), which form three dots when no voltage is applied on the substrate \cite{Pierre2009}. Table~\ref{tab:samples} describes the samples of each type (A, B and C) which are presented in the article. 
Fig.~\ref{fig1}i depicts samples used in Ref.~\onlinecite{Fujiwara2006} and \onlinecite{Angus2007} to build coupled dots in an intrinsic layer of silicon with the help of one large upper gate to attract electrons and two local top gates to create barriers.

\section{Effect of the doping profile on electron confinement}
\label{sec:LDD}

\begin{figure}[!b]
\begin{center}
\includegraphics[width=\columnwidth]{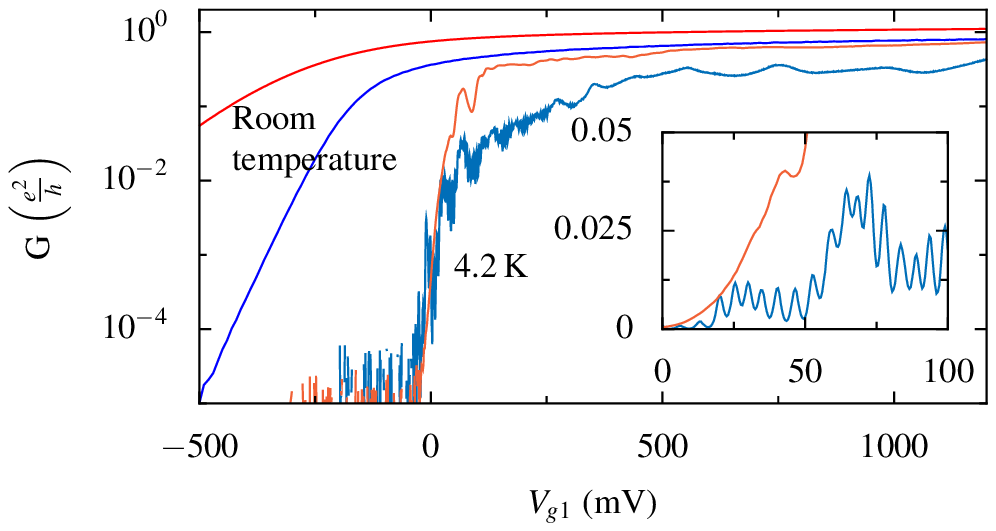}
\caption{Drain-source conductance of two double gate samples versus $V_{g1}$ at room temperature and 4.2\,K. $V_{g2}=1.2\,\mathrm{V}$, thus conductance is determined by the region below gate 1.
Blue curves: sample 1 (type A). Regular CBO are observed at 4.2\,K due to the formation of a  dot  below gate 1 and tunneling of electrons through the undoped nanowires located below the spacers \cite{Hofheinz06B,Pierre2009}. Red curves: sample 2 (type B). No CBO is observed. Inset:  detail at 4.2\,K in linear scale.}
\label{fig2}
\end{center}
\end{figure}

In this part we compare A- and B-type samples with no substrate bias applied.
In A-type samples, without LDD implantation, the undoped regions below the spacers become insulating at low temperature, resulting in the formation of a dot below the gate \cite{Hofheinz06B,Pierre2009}. In contrast confinement of carriers in the channel does not occur when the access regions are doped (B-type). This can be seen on Fig.~\ref{fig2} where the linear conductance of 2 two-gate samples, with and without LDD, is shown as a function of $V_{g1}$. Apart from LDD doping, these two samples are nominally identical. $V_{g2}$ is kept at a large positive value (1.2\,V) so that the conductance is controlled only by $V_{g1}$, the channel below gate 2 acting as a smaller, negligible and constant series resistance.
When the access regions are undoped clear and regular Coulomb blockade oscillations (CBO) are observed at 4.2\,K, i.e.\ a single-electron box is formed below gate 1. The period of these oscillations does not depend on the gate voltage and is 5\,mV, in agreement with what is expected for a MOS-SET with $L_g=W=60$\,nm, i.e.\ $\Delta V_{g} = \frac{e}{C_g}$ where $C_g$ is the gate-channel overlap geometrical capacitance \cite{Hofheinz06B,Pierre2009}. Only  aperiodic structure is observed  after LDD doping. Note that CBO are better seen just above threshold voltage because the tunnel barriers are progressively lifted when gate voltage $V_{g1}$ is increased.
The threshold voltage is close to -0.15\,V, the predicted value for an undoped Fully Depleted SOI MOSFET \cite{Poiroux}:
$$ V_{th}= -\frac{E_g}{2} + \frac{k_BT}{e} \ln \left(\frac{C_{ox}k_BT}{e^2n_iT_{Si}}\right)$$
where $E_g$ is the band gap of silicon, $n_{i}$ the intrinsic density of carriers, $T_{Si}$ the SOI thickness and $C_{ox} $ the gate oxide capacitance.
In both cases the aperiodic conductance fluctuation observed at low temperature, either modulated by CBO or not, is sample-specific and is due to the static disorder in the silicon nanowire below the spacers \cite{Hofheinz06}. This disorder is not screened  because of the low density of carriers in these regions.

The impact of LDD is also observed at room temperature. The ON-state conductance is higher with LDD doping, which proves that access resistances are lower in that case. We note that the conductance remains below $e^2/h$ in undoped samples, which is necessary to observe Coulomb blockade. 
LDD prevents confinement of carriers below a gate because it increases the access conductance above $e^2/h$.  
The sub-threshold slopes are also different. It is degraded in the case of LDD doping because the dopants diffuse below the gate, reducing the electrical gate length.

\section{Doped single dot obtained by LDD implantation}
\label{sec:singledotLDD}

\begin{figure}[!b]
\begin{center}
\includegraphics[width=1 \columnwidth]{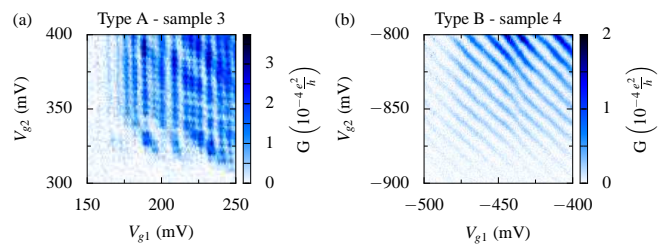}
\caption{Color plot of the drain-source conductance versus  $V_{g1}$ and $V_{g2}$ at 4.2\,K for two nominally identical double gate samples.
(a) Sample 3 (type A). Hexagons (barely seen at 4.2\,K, see also Ref.~\onlinecite{Pierre2009}) are observed above threshold.  (b) Sample 4 (type B). CBO are  periodic in ($V_{g1}+ V_{g2}$). Period of CBO is $10\,\mathrm{mV} \pm 1\,\mathrm{mV}$ and does not depend on gate voltages as expected for a strongly doped electron island located between the gates. The onset of the current is at negative gate voltages because of the shortening of the electrical gate length due to dopant diffusion.}
\label{fig3}
\end{center}
\end{figure}

When $V_{g2}$ is decreased to a value close to $V_{g1}$, A-type samples behave as double dot systems. Indeed a dot is formed below each gate whereas the undoped central region acts as a tunnel barrier (see our previous work \cite{Pierre2009}).
Fig.~\ref{fig3}a shows the corresponding hexagons and triple points close to the onset of current on both gates, at positive $V_{g1}$ and $V_{g2}$.
The honeycomb diagram is barely visible because first it is recorded at 4.2\,K (Data measured at 1\,K are shown in Ref.~\onlinecite{Pierre2009} on similar samples) and second, cotunneling is large. Indeed cotunneling lines are very sensitive to the interdot barrier transparency which can be large, sample specific, and not efficiently controlled by the two gates which do not directly cover the central region of the nanowire.
In strong contrast CBO oscillations are observed in samples with LDD (B-type).
They appear as periodic antidiagonal lines on Fig.~\ref{fig3}b where conductance is plotted as a function of $V_{g1}$ and $V_{g2}$. It means that a single dot is formed when both gate voltages $V_{g1}$ and $V_{g2}$ are close to their respective threshold. The period of these oscillations is approx.\ 10\,mV both for $V_{g1}$ and $V_{g2}$. Nearly equal periods means that the Coulomb island sits between the gates. This gives a coupling capacitance between this central dot and each gate of 16\,aF, in good agreement with an estimation based on the geometry. This capacitance is smaller than that between a MOS-SET and its covering gate (like in A-type samples), of  about 40\,aF \cite{Pierre2009}.
These oscillations are regular right from the onset of the current. This reflects the fact that the Coulomb island is  LDD doped and  contains a high density of electrons whatever $V_{g1}$ or $V_{g2}$, the one-particle mean level spacing being thus negligible.

\begin{figure}[!t]
  \begin{center}
 \includegraphics[width=1 \columnwidth]{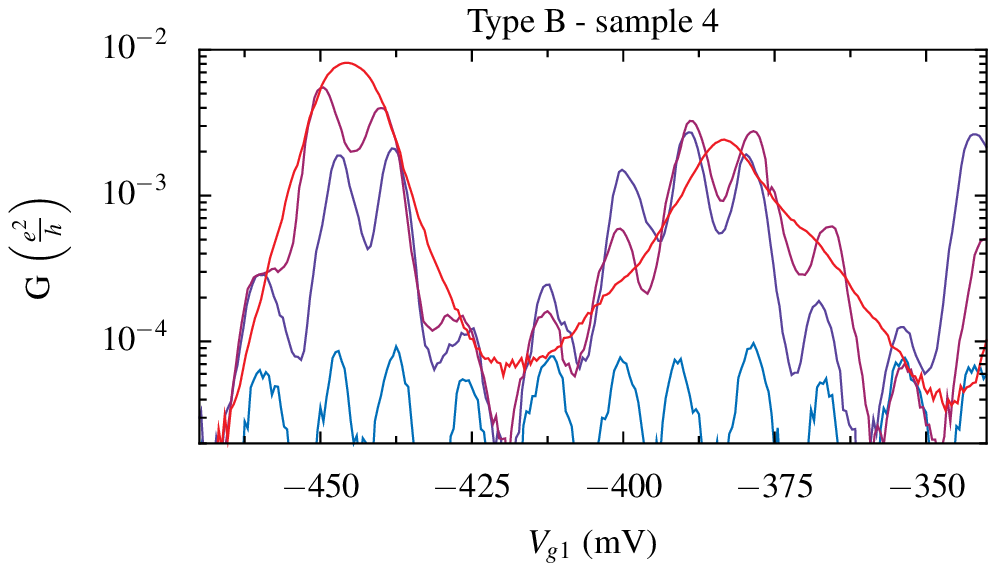}
  \caption{Drain-source conductance versus $V_{g1}$ at various $V_{g2}$ at 4.2\,K for sample 4. $V_{g2}=-700\,\mathrm{mV}$ (light blue), -550\,mV (dark blue), -150\,mV (violet), 250\,mV (red). For low $V_{g2}$ the current is limited mainly by barrier 2 tunnel conductance. At large $V_{g2}$  CBO are suppressed by  strong  coupling to the drain through barrier 2. The enveloping structure is due to gate modulation of the tunneling transmission under  gate 1. The shift towards positive $V_{g1}$  with decreasing $V_{g2}$  due to the cross talking between gates has been substracted.}
  \label{fig4}
  \end{center}
\end{figure}

The Coulomb island exists only if the conductance of the nanowire below each gate is small enough to confine electrons. Increasing one gate voltage (for instance $V_{g2}$ in Fig.~\ref{fig2}) suppresses the CBO. This is expected because the charge is not a good quantum number as soon as one of the two barriers has a conductance close to or above $e^2/h$. Fig.~\ref{fig4} shows how the CBO evolve with the increase of conductance of barrier 2. At $V_{g2}=-700\,\mathrm{mV}$ barrier 2 is more resistive than 1 and dominates the total conductance, leading to constant-height  oscillations. At $V_{g2}=-550\,\mathrm{mV}$  modulation of peak height arises from the dependence of the barrier 1 transparency with $V_{g1}$. The contrast of CBO is progressively reduced as the increased coupling to electrode 2 decreases the charging energy and favors co-tunneling ($V_{g2}=-150\,\mathrm{mV}$). At $V_{g2}=250\,\mathrm{mV}$ only the aperiodic pattern, similar to that shown on Fig.~\ref{fig2}, remains.
The possibility to tune  separately each tunnel barrier confirms our description of a dot sitting between the gates (Fig.~\ref{fig1}c).

\section{Silicon dots obtained by biasing the silicon substrate}
\label{sec:substrate_bias}

\begin{figure}[!t]
\begin{center}
\includegraphics[width=1 \columnwidth]{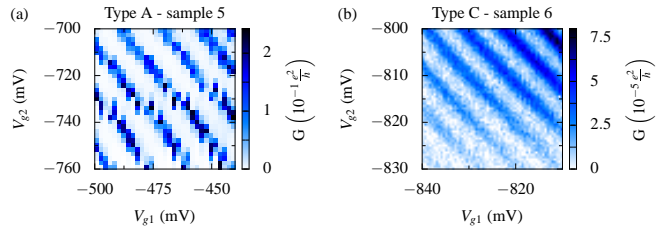}
\caption{Color plot of the drain-source conductance versus  $V_{g1}$ and $V_{g2}$ at 4.2\,K for two  samples without LDD. A strong positive voltage is applied on the substrate.
(a) Sample 5 with large spacers (type A), $V_{B}=20\,$V. The positive substrate bias increases the conductance of the intrinsic silicon as LDD doping does for B-type samples. At negative $V_{g1}$ and $V_{g2}$ a single dot and tunnel barriers separating it from source and drain are created. (b) Sample 6  with small spacers (type C), $ V_{B}=12.64\,$V. In both cases CBO are  periodic in ($V_{g1}+ V_{g2}$). The period of CBO is approx.\ twice as large in (a) than in (b) due to different sample width. The concentration of electrons in the central dot either by chemical (b) or electrostatic doping (a) is large. Thus the period of CBO is  independent on gate voltages.
Conductance is higher in (a) because it is measured well above threshold whereas in (b) it is measured just above threshold.}
\label{fig5}
\end{center}
\end{figure}

Both tunable  intrinsic and doped silicon quantum dots  can altenatively be formed by biasing the substrate without LDD doping. A single intrinsic dot is obtained in A-type samples. 
By applying a strong positive substrate bias (typically a few tens of volts) carriers are attracted from source and drain in the intrinsic silicon. This is the electrostatic analogue of the LDD chemical doping. Then strong negative voltages on $V_{g1}$ and $V_{g2}$ are needed to locally deplete the nanowire, in order to create a central dot isolated from the contacts. As in Ref.~\onlinecite{Fujiwara2006}, a fully tunable intrinsic silicon quantum dot is made. The CBO recorded at 4.2\,K are plotted on Fig.~\ref{fig5}a. They are  similar to the CBO observed in Fig.~\ref{fig3}b, except for the period in both $V_{g1}$ and $V_{g2}$ which is larger (approx.~18\,mV instead of 10\,mV), possibly indicating that the extension of the dot along the nanowire is smaller when confinement arises from electrostatics instead of doping modulation.

If small spacers are used (C-type samples), the doped central region acts as a single-electron box in series with the two MOS-SETs formed below the gates \cite{Pierre2009}. By biasing the substrate with a strong positive voltage, these two dots no longer exist because the access regions below the nitride spacers are made highly conductive. A tunable doped quantum dot is therefore obtained. 
As in the intrinsic version explained just before, the HDD implanted central dot is isolated from the source and drain by balancing  the positive backgate voltage with negative top gate voltages. 
The resulting CBO are shown at 4.2\,K on Fig.~\ref{fig5}b. The CBO are regular in gate voltage, like in Fig.~\ref{fig5}a and Fig.~\ref{fig3}b. The smaller period i.e.\ larger dot-gate capacitance can be explained by the 100\,nm width of this sample instead of $W=60$\,nm for the others .

There is no qualitative difference between the single intrinsic dot obtained by biasing the susbtrate and the single doped dots obtained both with LDD or HDD implantation between gates. This is due to the fact that chemical doping by implantation or electrostatic doping by gates are both large in our samples. LDD or HDD implantations induce high dopant and therefore high carrier concentration in our dots. In comparison the measured capacitance to substrate is approx.\ 20 times smaller than the capacitance to each top gate, such that a typical applied substrate bias of $+10$\,V  induces a surface carrier concentration of approx.\ 2\,$10^{12}\,\mathrm{cm}^{-2}$. The number of carriers is large even if the surface of our dots is very small (of the order of a few 1000\,nm$^2$) and quantum confiment effects which are very different for doped or intrinsic dots at small number of electrons are very small as compared to classical charging effects at these concentrations.

\section{Tunable implanted coupled dots}
\label{sec:tunable_double_dot}

\begin{figure}[!t]
\begin{center}
\includegraphics[width=1 \columnwidth]{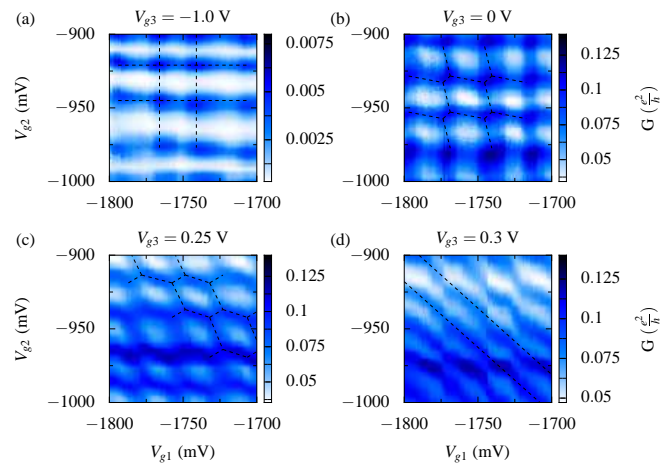}
\caption{Color plot of the drain-source conductance at 4.2\,K in sample 7 with three gates and with LDD implantation (type B). See Fig.~\ref{fig1}a.  The panels correspond to different voltages $V_{g3}$ applied to the central gate. $V_{g1}$ and $ V_{g2}$ are both strongly negative so that two dots are formed between gates 1 and 3 and gates 2 and 3. 
Dashed lines are guides for the eyes.
(a) Capacitance coupling between the two dots is negligible. The points at which both dots are non-blockaded are on a square lattice. The lines joining these points are due to cotunneling. (b) Inter-dot capacitive coupling is increased. The points on the square lattice are extended along the diagonal. (c) A clear honeycomb pattern is observed, characteristic of a strongly capacitively coupled double dot system. (d) The antidiagonal pattern of lines indicates the formation of a single large dot.}
\label{fig6}
\end{center}
\end{figure}

The technique of forming a single dot by implantation between top gates  can easily be extended to create coupled dots.
Samples with three gates (See Fig.~\ref{fig1}) behave as double quantum dots with tunable inter-dot coupling. Two dots can be created independently, one between gates 1 and 3, the other between gates 2 and 3. For the sample shown on Fig.~\ref{fig6} the three gates are smaller than in previous samples. Their length is 40\,nm, as can be seen on Fig.~\ref{fig1}. Moreover the two external gates partially cover the silicon nanowire, therefore greater negative gate voltages $V_{g1}$ and $V_{g2}$  have to be applied  to form tunnel barriers. 
Increasing $V_{g3}$ on the central gate permits to increase the  coupling  between the two dots.
Fig.~\ref{fig6} shows the evolution from two Coulomb islands with negligible capacitive coupling to a system with strong tunnel coupling \cite{Liu08}. The two dots eventually merge into a single larger dot spreading from gate 1 to gate 2 when the interdot tunnel conductance approaches $e^2/h$ \cite{Waugh95}. Contrarily to our double dot system obtained in undoped A-type samples, this structure enables to tune independantly the coupling between dots and their occupation.

As for single dots there is no qualitative difference between intrinsic and doped coupled dots, at least at the studied carrier concentration (For undoped A-type dots, the electron concentration is over $10^{12}\,\mathrm{cm}^{-2}$  when both gate voltages are above the threshold).
Our coupled As-implanted quantum dots have approximately the same size than their intrinsic silicon analogue \cite{Fujiwara2006,Pierre2009}.
The larger gate capacitances measured here (for example  $C_{g3}=24.2$\,aF  between the merged island   and central gate 3 instead of $C_{LGC}=6.7$\,aF in Ref.~\onlinecite{Fujiwara2006}) reflect   the smaller gate oxide thickness chosen in our case (5\,nm instead of 30\,nm). As in Ref.~\onlinecite{Fujiwara2006} our silicon dots can be treated within the orthodox theory where  quantum confinement effects are negligible. Therefore no difference is expected between doped and intrinsic dots. Quantum confinement effects are reported in Ref.~\onlinecite{Liu08} but at much lower temperature and lower carrier concentration.

\section{CONCLUSION}

In conclusion we have made fully tunable coupled silicon quantum dots by local implantation of As in a silicon nanowire.  
We have shown that combination of LDD  and HDD implantations with various self-aligned spacer lengths can be used to make versatile, very compact and tunable intrinsic or highly doped silicon quantum dots.
Local chemical doping borrowed from CMOS technology gives compact silicon dots with a high carrier number and excellent field-effect control of the tunnel barriers.   
The tunnel resistance of the barriers as well as the interdot coupling are fully controllable by top gates. These devices can become the building blocks of electron pumps, single electron adiabatic shuttle \cite{Skinner03} and, if scaled properly, silicon quantum bits \cite{Kane98}.

\section{Aknowledgement}
The research leading to these results has received funding from the European Community's seventh Framework (FP7 2007/2013) under the Grant Agreement Nr:214989. The samples subject of this work have been designed and made by the AFSID Project Partners http://www.afsid.eu.

\end{document}